# Scintillation counter with MRS APD light readout

A. Akindinov[a], G. Bondarenko[b], V. Golovin[c], E. Grigoriev[d], Yu. Grishuk[a], D. Mal'kevich[a], A. Martemiyanov[a], M. Ryabinin[a], A. Smirnitskiy[a], K. Voloshin[a,*]

[a]*Institute for Theoretical and Experimental Physics (ITEP), Moscow, Russia*
[b]*Moscow Engineering and Physics Institute, Moscow, Russia*
[c]*Center of Perspective Technologies and Apparatus (CPTA), Moscow, Russia*
[d]*University of Geneva, Switzerland*

**Abstract**

START, a high-efficiency and low-noise scintillation detector for ionizing particles, was developed for the purpose of creating a high-granular system for triggering cosmic muons. Scintillation light in START is detected by MRS APDs (Avalanche Photo-Diodes with Metal-Resistance-Semiconductor structure), operated in the Geiger mode, which have 1 mm$^2$ sensitive areas. START is assembled from a 15 × 15 × 1 cm$^3$ scintillating plastic plate, two MRS APDs and two pieces of wavelength-shifting optical fiber stacked in circular coils inside the plastic. The front-end electronic card is mounted directly on the detector. Tests with START have confirmed its operational consistency, over 99% efficiency of MIP registration and good homogeneity. START demonstrates a low intrinsic noise of about 10$^{-2}$ Hz. If these detectors are to be mass-produced, the cost of a mosaic array of STARTs is estimated at a moderate level of 2–3 kUSD/m$^2$.

## 1. Introduction

Scintillation counters are studied by the ITEP group in ALICE/LHC TOF project in the context of creating a cosmic triggering system, proposed for regular tests of a quantity of ALICE TOF modules in the course of their mass production and maintenance [1]. The objective is to build a cosmic ray telescope from a mosaic array of scintillating plates, which would cover several tens of m$^2$, provide 100% detection efficiency and which could be mounted onto a simple supporting structure.

A modern technique, which involves Avalanche Photo-Diodes (APDs) with a Metal-Resistance-Semiconductor (MRS) structure, operated in the Geiger mode [2], has been proposed for the registration of scintillation light. These detectors (hereinafter MRS APDs) were invented, designed and are currently produced at CPTA (Center of Perspective Technologies and Apparatus). Significant progress in the development of MRS APDs has been achieved over the last few years [3].

MRS APDs consume relatively little power and require a bias voltage of 50–60 V. They are sensitive to single photons at room temperatures and demonstrate intrinsic gain of up to 10$^6$ [4]. As compared to standard PMTs, MRS APDs do not require special housing or bulky light-conductors and can be mounted directly inside scintillating plates. This fact considerably simplifies the construction of a large-scale triggering system.

## 2. START construction

Scintillator-fiber systems represent a popular tool for particle registration, and they have been implemented in a number of experiments in high-energy physics. As an example, this technique has been in use for several years by the HERA-B electromagnetic calorimeter at DESY, the scintillation light there being detected by PMTs [5].

---

[*] Corresponding author.
*E-mail address:* `Kirill.Voloshin@itep.ru` (K. Voloshin).

START (as abbreviated from **S**cintillation **T**ile with MRS **A**PD Light **R**eadou**T**) is a detector of ionizing particles, assembled from a scintillating plate, two MRS APDs and two pieces of wavelength-shifting (WLS) optical fibers. Its key distinction from other detectors of this kind lies in the fact that START uses high-gain MRS APDs as photosensors. The scintillator size, as dictated by geometry of the ALICE TOF test facility, has been set equal to $150 \times 150 \times 10$ mm$^3$. Several types of plastic scintillators were tested, concentrating mainly on BC-412[1]. To get the best internal reflection and light collection, all the plates were diamond polished and wrapped in an opaque material.

The influence of MRS APD intrinsic noise on detector performance has been significantly decreased by fitting out one scintillating plate with two MRS APDs, plugged into a coincidence gate. The shaping time is usually set in the range of 20–80 ns, providing noise reduction of several orders of magnitude.

The appearance of START is shown in Fig. 1. The collection of light inside the plastic and its transportation towards MRS APDs is performed by commonly available WLS fibers, 1 mm in diameter. Fiber flexibility allows for packing them in several circular coils, thus increasing the light collection efficiency over the plastic volume. One end of each piece of fiber is covered with reflecting foil, while the other is pressed to the sensitive surface of MRS APD. Two to three fiber coils collect a sufficient amount of light. To simplify the construction of the detector, the fibers are stacked in narrow circular grooves, engraved in the plastic. The ring diameter is chosen close to the external dimensions of the detector.

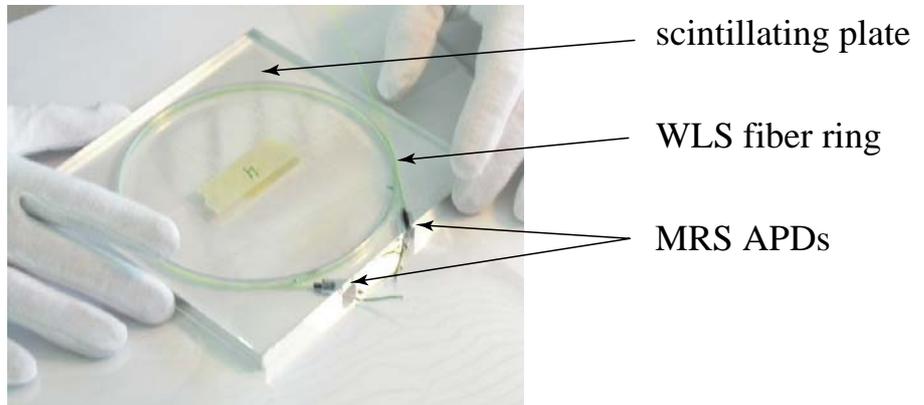

Fig. 1. START assemblage: scintillating plastic plate, WLS fibers and two MRS APDs put together.

## 3. MRS APD

APDs have long since been invented and integrated into scintillator-fiber systems for the needs of particle physics, astrophysics and medical imaging. APDs demonstrate fast response and high quantum efficiencies, and they are under intensive study in the framework of design and construction of LHC experiments [6]. Radiation-hard types of APDs have been developed to be used on a large scale in the CMS electromagnetic calorimeter [7]. The relatively low gain (~$10^3$) of APD is its main disadvantage, restricting its use for light-detection [8].

The drastic increase in APD gain can be achieved by setting the bias voltage above the breakdown point. This mode of operation (the Geiger mode) is implemented in MRS APD, which represents a 1 mm$^2$ matrix of micro-cells with MRS intrinsic structure, each sized $20 \times 30$ μm$^2$, processed on a common p-type silicon substrate, as illustrated in Fig. 2. To decrease the overall electric resistance, the substrate is heavily doped with p$^+$ from the bottom. The n$^+$-doping of micro-cells creates a high electric field in the vicinity of the p–n-junction. This field is sufficient for the development of local avalanches from the initial photo-ionization in exposed micro-cells. Avalanche quenching is ensured by a voltage drop on the resistive layer, which is located under the upper contact.

---

[1] Produced by Saint-Gobain Crystals & Detectors, www.bicron.com.

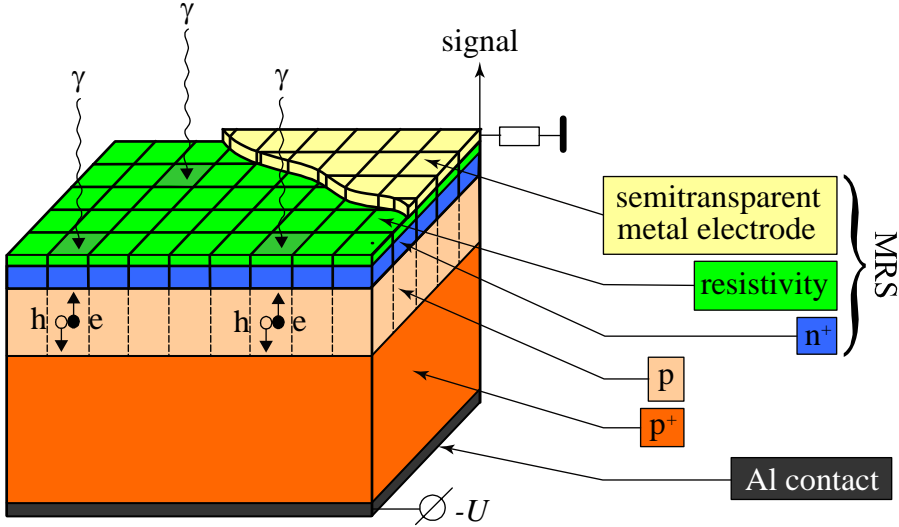

Fig. 2. General look of MRS APD construction.

Intensive studies of MRS APDs as possible photosensors for scintillation detectors, carried out by the ITEP group in ALICE in 2003, gave encouraging results [4]. The spectral sensitivity of MRS APD is shifted in respect to the spectrum of scintillation light towards longer waves, being as high as 35% for the green light (520 nm) [9]. This fact motivates the use of either more 'red' scintillating plastic or a wavelength-shifting technique. The latter idea was realized in START by means of WLS fibers re-emitting the light in the region close to 500 nm. Experimenting with different fiber types has resulted in a multi-fold increase of the MRS APD output. Double cladding Kuraray Y11[2] fibers have been finally chosen as the optimal solution for START.

The MRS APD gain can possibly be as high as $10^6$. In reality, the optimal working point corresponds to the gain value set between $10^5$ and $10^6$. If the discriminating threshold is fixed at 3–4 photo-electrons, the noise rate of MRS APD is about $10^3$–$10^4$ Hz. In this case, setting the coincidence gate length for two MRS APDs in START at 10–20 ns reduces the overall detector noise to ~$10^{-2}$ Hz, which is a negligible quantity. A detailed discussion of the MRS APD parameters may be found in Ref. [4].

## 4. Detector performance

A special front-end electronic (FEE) card for START signal processing was designed, tested and built. Since a single detector contains two MRS APDs, the card involves two separate amplifying channels and two resistive dividers for independent setting of the bias voltage in the range 50–60 V. The following averaged parameters of the card have been measured:

(1) Amplification — 0.8 V/pC.
(2) Noise ($C_{input} = 0$) — less than 1500 electrons.
(3) Linearity — up to 300 fC at the input level.
(4) Timing resolution of the coincidence circuit — 20 ns.

A detailed description of the FEE card may be found in Ref. [10]. The card is compact and may be attached directly to the START body as shown in Fig. 3. Contacts of MRS APDs are solded to the card with no additional wiring.

Multiple beam and cosmic tests were carried out at ITEP and CERN in 2003. Various START options and characteristics were studied, including plastic, fiber and APD types, detector geometry, light collection, etc.

First START samples were studied by means of pion and proton beams with several GeV/$c$ momenta. The trigger was generated by scintillation counters positioned along the beam path. To clarify the pedestal position, amplitude measurements were performed with a wide beam, so that some particles missed the detector. Fig. 4 represents a typical amplitude distribution from one of the two

---

[2] Produced by Kuraray Co., Ltd., www.kuraray.co.jp/en/index.html.

MRS APDs mounted in START obtained under these conditions. A clear gap separates the trigger events from the pedestal.

The number of photo-electrons can be obtained in three different ways as comprehensively discussed in Ref. [4]. Since the START construction has been fixed, the number of photo-electrons depends only on the type of plastic. A value measured for plates of BC-412 and Polisterol 165[3] equals to 14–17 electrons/MIP. The MIP registration efficiency in these cases was found to be more than 99%.

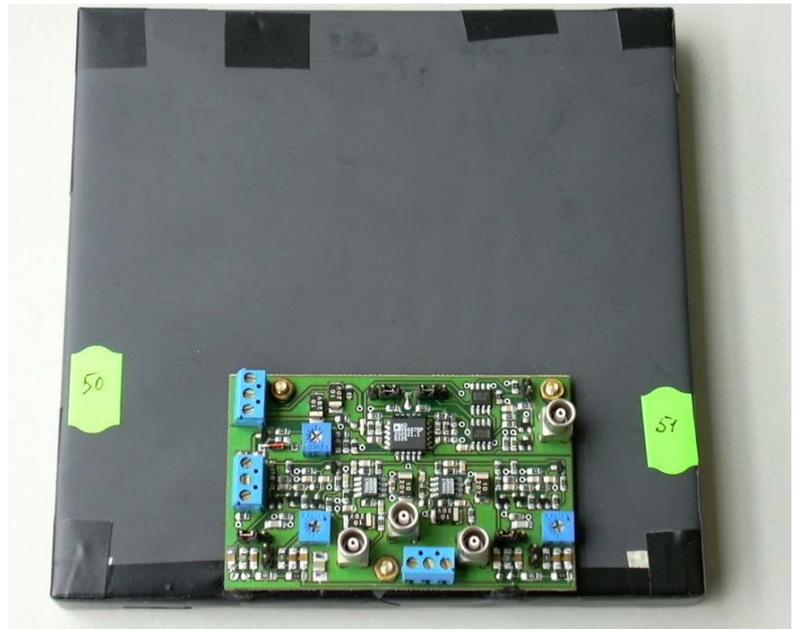

Fig. 3. Assembled START with the FEE card.

The discriminator thresholds are usually set in the range 80–120 mV. The cosmic triggering system applies mild restrictions on how precisely the working point of START can be chosen. The bias voltage may be set with an accuracy of 1–2%, the discriminator threshold with an accuracy of 10%.

Surface uniformity of START was thoroughly scanned with a small-size beam. Amplitude characteristics in the most important points (i.e. in the center, outside the fiber ring, at some point exactly on the fiber ring and at the edge) are shown in Fig. 5. When the beam pointed at the edge, some particles missed the detector, making the pedestal position clear in the corresponding spectrum (bottom left). It may be seen that the four spectra look much alike, the mean values being the same within 8–10% of accuracy. The efficiency remains above 99%. These factors prove that START is sufficiently homogeneous and is an appropriate solution for the cosmic triggering system.

## 5. Cost estimates

The cost of a single detector can be roughly estimated at 70–140 USD, half of which is the cost of MRS APDs. This value is likely to be closer to 70 USD under mass production. Covering large areas with STARTs will require 30–40 detectors/m$^2$ and should cost as low as 2–3 kUSD/m$^2$. The development of technology over the last two years has already resulted in a 3–4-fold decrease of the MRS APD cost. The prediction is that this cost reduction will continue. Moreover, the FEE card may get cheaper if a more integrated design is employed.

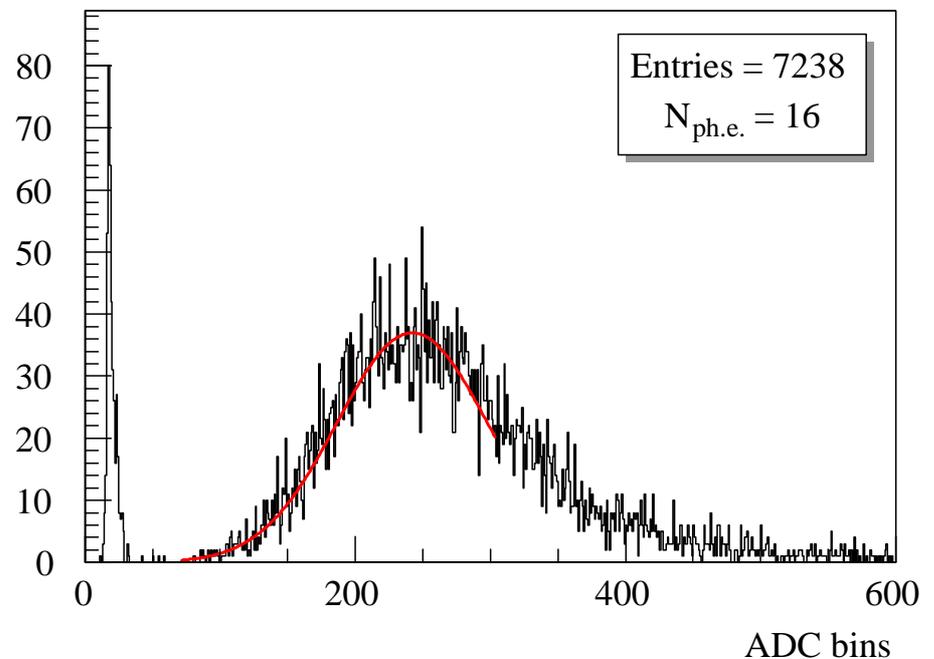

Fig. 4. Amplitude spectrum from one of the two MRS APDs mounted in START, measured with a wide pion beam. The pedestal is formed by events in which particles do not hit the detector.

---

[3] Produced by Polimersintez, B. Nizhegorodskaya 77, Vladimir, 600016, Russia.

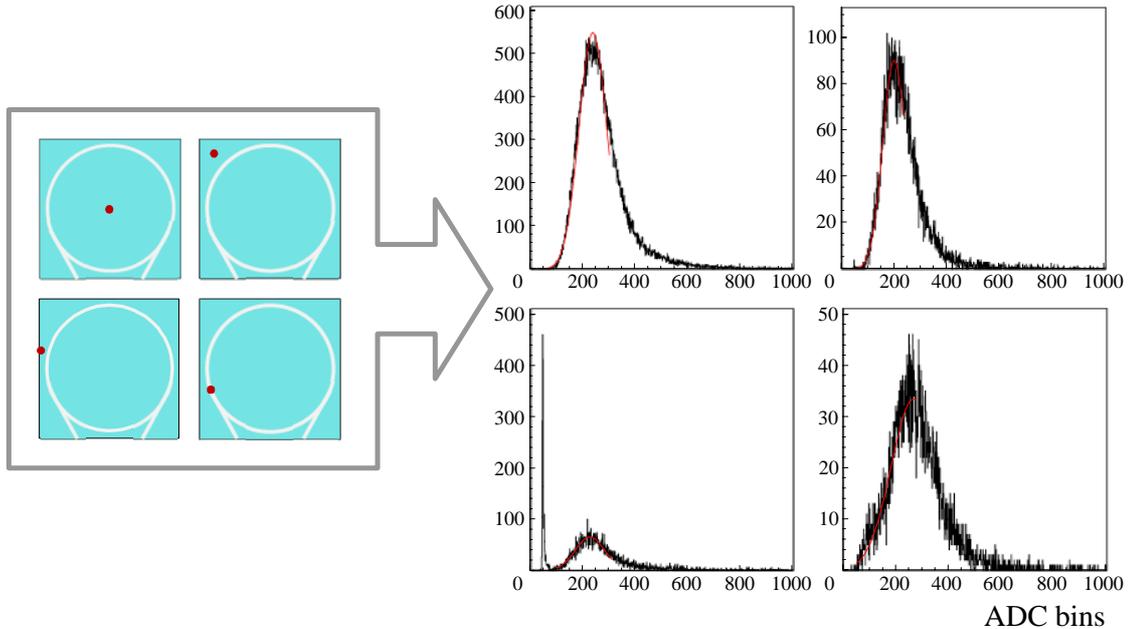

Fig. 5. START amplitude spectra measured by a beam hitting various parts of the detector surface: the center, a point outside the fiber ring, a point exactly on the fiber ring and a point at the edge.

## 6. Conclusion

Scintillating plates with MRS APD light readout show stable characteristics. Signals produced in response to MIP are clear and stay far above the pedestal value. Registration efficiency remains close to 100% over the whole detector surface. Required voltage supply is about 50 V. Intrinsic noise stays at the level of $10^{-2}$ Hz and can be easily decreased to $10^{-3}$–$10^{-4}$ Hz. Due to their geometry, STARTs can be arranged in large arrays to form a high-granular low-noise detecting system.

We would like to express our gratitude to M. Danilov, V. Rusinov, V. Sheinkman and E. Tarkovskiy for their valuable help in carrying out this research.

## Acknowledgements

We would like to express our gratitude to M. Danilov, V. Rusinov, V. Sheinkman and E. Tarkovskiy for their valuable help in carrying out this research.